\documentclass[letterpaper,twocolumn,showpacs,amsfonts,aps,%
                             superscriptaddress,prc,nofootinbib,floatfix]{revtex4}

\usepackage{fullpage}
\usepackage{amsmath}
\usepackage{bm}
\usepackage{graphicx}

\usepackage{color}
\usepackage[colorlinks=true, linkcolor=black, citecolor=blue, urlcolor=blue]{hyperref}

\newcommand{\ecc}{\varepsilon}

\newcommand{\snn}{\sigma_\mathrm{NN}^\mathrm{inel}}
\newcommand{\sgg}{\sigma_{gg}}
\newcommand{\Npart}{N_\mathrm{part}}
\newcommand{\beq}[1]{\begin{equation}\label{#1}}
\newcommand{\eeq}{\end{equation}}

\newcommand{\ie}{\emph{i.e. }}

\newcommand{\eq}{{\,=\,}}

\newcommand{\rperp}{\bm{r}_\perp}

\newcommand{\bb}{\bm{b}}
\newcommand{\br}{\bm{r}}

\def\La{\langle}
\def\Ra{\rangle}

\begin{document}


\title{Energy dependent growth of the nucleon and hydrodynamic initial conditions}

\author{Ulrich Heinz}
\email[Correspond to\ ]{heinz@mps.ohio-state.edu}
\affiliation{Department of Physics, The Ohio State University, 
Columbus, OH 43210, USA}
\author{J. Scott Moreland}
\affiliation{Department of Physics, The Ohio State University, 
Columbus, OH 43210, USA}
\affiliation{Institut f\"ur Theoretische Physik, Johann~Wolfgang~Goethe-Universit\"at,
                  Max-von-Laue-Stra\ss e 1, D-60438 Frankfurt am Main, Germany}

\begin{abstract}
Due to gluon saturation, the growth of the inelastic nucleon-nucleon cross section 
with increasing collision energy $\sqrt{s}$ results in a broadening of the nucleon's 
density distribution in position space. This leads to a natural smoothing of the
initial energy density distribution in the transverse plane of the matter created near 
midrapidity in heavy-ion collisions. We study this effect for fluctuating initial conditions
generated with the Monte Carlo Kharzeev-Levin-Nardi (MC-KLN) model for Au+Au
collisions at the Relativistic Heavy Ion Collider (RHIC) and the 
Large Hadron Collider (LHC). We argue that at the LHC viscous hydrodynamics is
applicable at earlier times than at RHIC, not only because of the higher temperature
but also since the length scale over which the initial pressure fluctuates increases
with collision energy. 
\end{abstract}

\date{\today}

\maketitle

\section{Introduction}
\label{sec1}

Hydrodynamics has been successful in modeling the hot, dense matter produced in 
ultra-relativistic heavy-ion collisions, particularly in describing the large collective flow 
observed at RHIC and the LHC \cite{Kolb:2003dz}. A fully quantitative description of the
experimental data requires, among other ingredients, realistic initial conditions. These 
initial conditions currently remain a significant source of uncertainty in predicting final 
state obervables.

Optical models \cite{Glauber,Kolb:2001qz,KLN} for the 2-dimensional initial density 
profile in the plane transverse to the beam axis, generated from the overlap of 
longitudinally integrated Woods-Saxon distributions ("nuclear thickness functions"), 
yield radially symmetric entropy and energy densities for zero impact parameter 
($b\eq0$) collisions. This symmetry drives radially symmetric hydrodynamic expansion, 
with zero anisotropic flow coefficients $v_n{\,\equiv\,}\La\cos(n\phi)\Ra$ (where $\phi$ 
is the azimuthal angle in the transverse plane relative to the impact parameter, and the 
average is taken with the azimuthal distribution of the final state particles). The 
non-vanishing elliptic flow $v_2$ observed in central ($b{\,\approx\,}0$) Cu+Cu and 
Au+Au collisions at RHIC \cite{Alver} and the non-zero harmonic flow coefficients
$v_1,\dots,v_6$ measured in central Pb+Pb collisions at the LHC
\cite{:2011vk,CMSflow,Steinberg:2011dj} are in direct conflict with this prediction and 
require new physics to model the density of matter produced in relativistic heavy-ion
collisions. 

Collective flow is the hydrodynamical response to the pressure gradients in an individual 
collision event, and the latter fluctuate from event to event. In essence, each nuclear 
collision makes a quantum mechanical measurement of the actual transverse nucleon 
positions inside each nucleus at the time of collision, and the outcome of this measurement fluctuates stochastically even though the corresponding probability distribution (the average nuclear density profile) is smooth. Therefore, the initial density distribution of matter 
produced even in a perfectly central ($b\eq0$) collision is {\it not} azimuthally symmetric 
(and the resulting anisotropic flow coefficients $v_n$ do {\it not} vanish) even though the 
ensemble-averaged initial density has this symmetry.

This was first pointed out by Miller and Snellings \cite{Miller:2003kd} who demonstrated
the importance of event-by-event initial state fluctuations on anisotropic collective flow 
using a Monte Carlo implementation of the Glauber model \cite{Miller:2007ri}. A 
Monte Carlo version of the KLN model \cite{KLN}, which incorporates the idea of gluon
saturation at high collision energies, was later developed by Drescher and Nara
\cite{Drescher:2006ca} and further improved in 
Refs.~\cite{Hirano:2009ah,Albacete:2010ad,Albacete:2011fw}.

The original MC-KLN code \cite{Drescher:2006ca,Hirano:2009ah} modeled nucleons
as homogeneous cylinders along the beam direction whose cross section is determined 
by the (energy dependent) inelastic nucleon-nucleon scattering cross section $\snn$. This 
leads to step-function-like discontinuities in the nuclear thickness function $T_A(\rperp)$, 
which (through the gluon saturation momentum $Q_s^2(\rperp)$ \cite{KLN}) controls the 
initial transverse entropy density profile $s(\rperp)$ of the collision fireball \cite{Hirano:2005xf}. 
These artificial discontinuities generate big fluctuations in the initial hydrodynamic velocity 
gradients which create such large viscous pressure gradients that viscous hydrodynamics 
breaks down. This can be avoided by using a more realistic Gaussian profile for the 
nucleon density \cite{Albacete:2011fw}. In \cite{Albacete:2011fw} it was assumed that 
small-$x$ evolution is local in impact parameter space, which may underestimate the
broadening of the dipole scattering amplitude in impact parameter space with increasing
collision energy. We here point out that, if one follows the idea of gluon saturation, the width 
of the gluon distribution inside a nucleon should grow with energy, due to transverse diffusion 
of small-$x$ gluons: as $\sqrt{s}$ increases, both $\snn$ and $\sigma$ should become 
larger.  We model this by an energy-dependent broadening of the Gaussian nucleon thickness
function $T_p(\bm{r}_\perp)$.\footnote{The possibility of measuring this broadening at a 
  future electron-ion collider has recently been pointed out in Ref.~\cite{Horowitz:2011jx}.
  We also note that similar ideas precede the development of the concept of gluon
  saturation and follow from Froissart's work on the unitarization of scattering amplitudes
  \cite{Froissart:1961ux}, see \cite{Martin} for a review.}   
We show that this broadening leads to a smoothing of fluctuations in the initial density 
distribution of higher-energy heavy-ion collisions and to a corresponding decrease of 
the higher order eccentricity coefficients $\ecc_{n>2}$ (which drive the higher order 
anisotropic flow coefficients $v_{n>2}$ \cite{Qiu:2011iv}) at the LHC compared to RHIC.  

\section{Sizing the nucleon}
\label{sec2}

We consider Gaussian nucleons with the normalized density distribution
(the subscript $p$ stands for "proton")
\begin{equation}
  \label{eq1}
  \rho_p(\br)=\frac{e^{-r^2/(2B)}}{(2\pi B)^{3/2}},
  \qquad \left(\textstyle{\int d^3r \, \rho_p(\br) = 1}\right), 
\end{equation}
corresponding to a 3-dimensional rms radius of the nucleon of 
$r_p^\mathrm{rms}\eq\sqrt{\La r^2\Ra}\eq\sqrt{3B}$. 
Eq.~(\ref{eq1}) yields the nucleon thickness function (defined as the integral
of the nucleon density along the beam axis)
\begin{eqnarray}
\label{eq2}
  && T_p(\rperp) = \int dz\, \rho_p(\rperp,z) 
                              = \frac{e^{-r^2/(2B)}}{2\pi B},
\nonumber\\
  && \int d^2r_\perp\, T_p(\rperp) = 1, 
\end{eqnarray}
corresponding to a transverse rms radius $r_\perp^\mathrm{rms}\eq\sqrt{2B}$.
We interpret it as the longitudinally integrated (``transverse'') density of colored 
partons (``gluons") inside the proton. 

In this section, we determine the Gaussian width $B$ as a function of $\sqrt{s}$
by relating it to the $\sqrt{s}$-dependent inelastic nucleon-nucleon cross section 
$\snn$. Motivated by the Glauber-Mueller formula for the impact parameter dependent
cross section of a color dipole with the quarks and gluons inside a proton 
\cite{Mueller:1989st,Kowalski:2003hm}, we make a Glauber-like ansatz 
\cite{Albacete:2011fw} for the probability $P(b)$ of a nucleon-nucleon collision
at impact parameter $b$:
\begin{equation}
\label{eq3}
  P(b) = 1 - \exp[-\sgg T_{pp}(b)].
\end{equation}
Here $T_{pp}(b)$ is the transverse density for binary collisions between gluons, 
\begin{eqnarray}
\label{eq4}
  && T_{pp}(b) = \int d^2r_\perp\, T_p(\rperp)\, T_p(\bb{-}\rperp)
                    = \frac{e^{-b^2/(4B)}}{4\pi B},
                    \nonumber\\
  && \int d^2b\, T_{pp}(b) = 1,
\end{eqnarray}
and $\sgg$ is the glue-glue interaction cross section.

We are mostly interested in the initial transverse density distribution 
of the matter created at mid-rapidity in heavy-ion collisions. As the
collision energy increases, the production processes for this matter
involve gluons of smaller and smaller longitudinal momentum 
fraction $x$. As $x$ decreases, the gluon density (gluon distribution 
function) grows \cite{Deshpande:2005wd}. At high gluon densities, 
gluons start to recombine \cite{Mueller:1985wy}, leading eventually 
to a saturation of the gluon density \cite{JalilianMarian:2005jf}.
For a Gaussian density distribution as in Eq.~(\ref{eq1}) gluon saturation
becomes effective in the center of the proton first, reaching into the
periphery only at still higher collision energies. This leads to a 
flattening of the normalized density distribution (\ref{eq1}) near its
center, accompanied by a growth of its rms radius. Gluon saturation
effects can therefore be mimicked by allowing the parameter $B$ in
Eq.~(\ref{eq1}) to grow with $\sqrt{s}$. This is what we suggest in the
present article; it differs from the treatments in 
\cite{Kowalski:2003hm,Albacete:2011fw} which held $B$ constant.

The inelastic nucleon-nucleon cross section $\snn$ is obtained by 
integrating the inelastic gluon-scattering probability $P(b)$ in 
Eq.~(\ref{eq3}) over all impact parameters:
\begin{equation}
\label{eq5}
  \snn = \!\!\int\!\! d^2b\, P(b)
           = \!\!\int\!\! d^2b\, \Bigl(1 - \exp[-\sgg T_{pp}(b)]\Bigr).
\end{equation}
In our model the $\sqrt{s}$ dependence of $\snn$ results from the combined
energy dependences of $\sgg$ and the Gaussian width $B$ of $T_{pp}(b)$. 
This differs from the work \cite{Albacete:2011fw} where the $\sqrt{s}$
dependence of $\snn$ arises entirely from the energy dependence of $\sgg$.

With the explicit form (\ref{eq4}) for the binary gluon collision density and the
substitutions $t\eq{b^2}/(4B)$ and $\lambda\eq\sgg/(4\pi B)$, relation 
(\ref{eq5}) can be written as
\begin{equation}
\label{eq6}
  \frac{\snn}{4\pi B} = f(\lambda),
\end{equation}
where
\begin{equation}
\label{eq7}
  f(\lambda) = \int_0^\infty dt\, \Bigl(1- \exp(-\lambda\,e^{-t})\Bigr)
\end{equation}
is a monotonically increasing function with $f(0)\eq0$.

To completely specify our model we assume that $\lambda\eq\sgg/(4\pi B)$
is an energy independent constant, or (equivalently) that the Gaussian width $B$ 
(which is proportional to the square of the proton radius) increases with collision
energy proportionally to the inelastic nucleon-nucleon cross section $\snn$.
This assumption makes intuitive sense: as the collision energy increases,
wee partons in the tail of the proton's density distribution get boosted to
sufficient energy to contribute to the scattering between nucleons while a
similar increase in the dense center of the proton gets tempered by gluon
saturation. As a result, the proton gets bigger. The glue-glue interaction cross 
section $\sgg$ increases at the same rate; in fact, since $P(b)$ is always smaller
than 1 (due to the non-vanishing "punch-through" probability described
by the last term in Eq.~(\ref{eq3})), $\sgg{\,>\,}\snn$ is always true.\footnote{There 
  is no fundamental reason for assuming $\lambda = \sgg/(4\pi B)$ to be constant.
  In fact, asymptotic freedom in QCD suggests that $\sgg$ might grow more slowly 
  with $\sqrt{s}$ than $B$, in which case $\lambda$ and $f(\lambda)$ would be 
  decreasing functions of $\sqrt{s}$, and the proton size $B$ would grow more strongly 
  with energy than the inelastic nucleon-nucleon cross section. This would further 
  increase the smoothing effects on initial state fluctuations at higher collision energies 
  to be discussed below.}  

To fix $\lambda$ we use experimental input at fixed collision energy
that is sufficiently large for the Glauber model \cite{Glauber} and our
gluon-saturation-based ideas for extrapolating to higher energies to 
work, but small enough that we can still make contact with the measured
and tabulated nuclear profiles at low energies, without having to
take into account energy-dependent swelling of the entire nucleus.

At $\sqrt{s}\eq23.5$\,GeV, the inelastic nucleon-nucleon cross section
has been measured as $\snn\eq3.2$\,fm$^2$ \cite{Nakamura:2010zzi}.
An extraction of the rms impact parameter $b_\mathrm{rms}\eq\La b^2\Ra^{1/2}$ 
for inelastic proton-proton collisions from CERN ISR data has been 
reported by Amaldi and Schubert \cite{Amaldi:1979kd}: at 
$\sqrt{s}\eq23.5$\,GeV they find $b_\mathrm{rms}\eq1.03$\, fm.
Our model allows to compute $b_\mathrm{rms}$ from the collision
probability $P(b)$ in Eq.~(\ref{eq3}) as follows:
\begin{eqnarray}
\label{eq8}
  \La b^2\Ra &=& \frac{\int d^2b\, b^2\, P(b)}{\int d^2b\, P(b)}
\\
  &=& 4B\,\frac{\int_0^\infty dt\,t\,(1-\exp(-\lambda e^{-t}))}
                         {\int_0^\infty dt\,(1-\exp(-\lambda e^{-t}))}
  \equiv 4B\,g(\lambda).
\nonumber
\end{eqnarray}
Equations (\ref{eq6}) and (\ref{eq8}) can be cast into the form
\begin{equation}
\label{eq9}
  B = \frac{\snn}{4\pi f(\lambda)} 
      = \frac{\La b^2\Ra}{4 g(\lambda)},
\end{equation}
which can be solved for both the energy-independent constant 
$\lambda$ and the proton width parameter $B$ at 
$\sqrt{s}\eq23.5$\,GeV. We find $\lambda\eq1.62$, corresponding 
to $B(23.5\,\mathrm{GeV})\eq(0.473\,\mathrm{fm})^2$. At other
collision energies $B$ can now be computed from
\begin{equation}
\label{eq10}
  B\left(\sqrt{s}\right) = \frac{\snn(\sqrt{s})}{14.30\,\mathrm{fm}^2}.
\end{equation}
In Table~\ref{T1} we collect a few representative values.

%
\begin{table}[ht!] 
\caption{The Gaussian width $\sqrt{B}$ (Eq.~(\ref{eq10}))
   for various collision energies. The values for $\snn$ at LHC energies
   (2.76 and 7\,TeV) were reported in \cite{Collaboration:2011rta,Aad:2011eu}.
   \label{T1}}
\label{tab:1}   
\begin{center}
\begin{tabular}{ccccc}
  \hline\hline
  \ \ $\sqrt{s}$\ (GeV) &\vline & $\snn$\ (mb) &\vline & $\sqrt{B}$\ (fm)\ \  \\
  \hline
  23.5  	&\vline & 32	&\vline	 & 0.473 \\
  200		&\vline & 42	&\vline	 & 0.544 \\	
  2760	&\vline & 62.1	&\vline	 & 0.661 \\
  7000 	&\vline & 71	&\vline	 & 0.707 \\
  \hline\hline
\end{tabular}
\end{center}
\end{table}
%

Before describing how we distribute these ``growing nucleons'' inside the colliding 
nuclei, let us summarize the novel features of our approach. In \cite{Albacete:2011fw}
all of the energy dependence of $\snn$ on the l.h.s. of Eq.~(\ref{eq5}) was attributed
to the factor $\sgg$ in the exponent on the r.h.s. of Eq.~(\ref{eq5}). Here we split this
energy dependence between the factors $\sgg$ and $T_{pp}(b)$ on the r.h.s. of
Eq.~(\ref{eq5}), by letting {\em both} $\sgg$ and the nucleon size $B$ grow with 
$\sqrt{s}$. Our specific model assumption (for which we have no convincing argument
other than simplicity) is that all three quantities ($\snn$, $\sgg$ and $B$) grow at exactly 
the same rate. While this may not be completely correct, {\em some} growth of the 
nucleon size $B$ with $\sqrt{s}$ is suggested by theory \cite{Gotsman:2010nw} and 
required by experimental data \cite{Amaldi:1979kd,Gotsman:2010nw,TOTEM}: 
Amaldi and Schubert demonstrated a moderate growth of the proton {\it inelastic} 
interaction radius over the energy range probed by the ISR \cite{Amaldi:1979kd}, 
and Fig.~19(E) in Ref.~\cite{Gotsman:2010nw} together with the recent TOTEM 
result from elastic $pp$ scattering at $\sqrt{s}\eq7$\,TeV at the LHC provide 
experimental evidence for a significant growth of the proton's {\em elastic} interaction
radius by a factor $\sim2$ between low ISR and top LHC energies, roughly comparable
with the similar increase of $\snn$ over the same energy range documented in
Table~\ref{T1}. It is this growth of the proton radius that causes the smearing of
event-by-event fluctuations in very high energy nuclear collisions to be discussed in
the following.

\section{Distribution of nucleons}
\label{sec3}

\begin{figure}[hbt]
  \includegraphics[width=\linewidth]{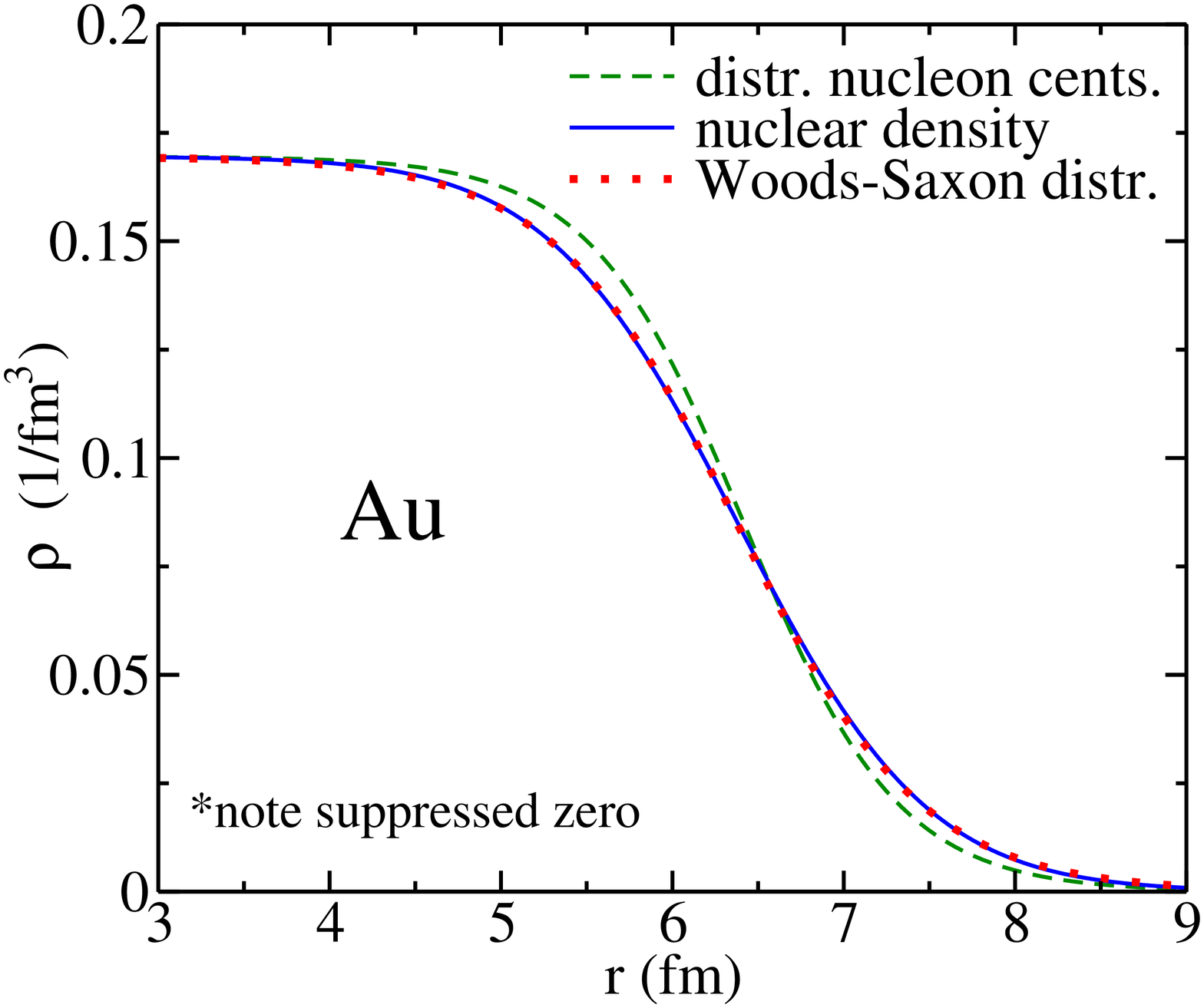}
  \caption{(Color online) Nuclear density $\rho_\mathrm{Au}(r)$ (Eq.~(\ref{eq11}), solid line)
  and the distribution of nucleon centers $\tilde\rho_\mathrm{Au}(r)$ (Eqs.~\ref{eq12},\ref{eq14}),
  dashed line), together with the Woods-Saxon distribution (\ref{eq12}) with measured
  parameters (\ref{eq14}) (dotted line). Please note that only the tail of the distribution
  is plotted, to better see the differences between $\rho_\mathrm{Au}(r)$ and
  $\tilde\rho_\mathrm{Au}(r)$.     
  \label{F1}
  }
\end{figure}

To describe the two incoming nuclei before the collision, we distribute the nucleons 
(\ref{eq1}) with a distribution $\tilde{\rho}_A(\br_0)$ (where $\br_0$ denotes the center 
of the nucleon) that is adjusted such that the resulting folded distribution
\cite{Hirano:2009ah}
\begin{equation}
\label{eq11}
  \rho_A(\br) = \int d^3r_0\, \tilde{\rho}_A(\br_0)\,\rho_p(\br{-}\br_0)
\end{equation}
reproduces the measured nuclear density profile for a nucleus of mass number $A$ 
\cite{De Jager:1987qc}. We take both $\rho_A$ and $\tilde{\rho}_A$ to be of 
Woods-Saxon form:
\begin{eqnarray}
\label{eq12}
  \rho_A(r) &=& \frac{\rho_0}{\exp\left(\frac{r{-}R_A)}{d_A}\right) + 1},
\\
\label{eq13}
  \tilde\rho_A(r_0) &=& 
  \frac{\tilde\rho_0}{\exp\left(\frac{r_0{-}\tilde{R}_A)}{\tilde{d}_A}\right) + 1}.
\end{eqnarray}
The measured values for gold are \cite{De Jager:1987qc,Hirano:2009ah}
\begin{equation}
\label{eq14}
  \rho_0\eq0.1695\,\mathrm{fm}^{-3},\ 
  R_\mathrm{Au}\eq6.38\,\mathrm{fm},\ 
  d_\mathrm{Au}\eq0.535\,\mathrm{fm}.
\end{equation}
We find that the choice $\tilde\rho_0\eq\rho_0$ works with excellent accuracy
(see Fig.~\ref{F1}). We perform a two-parameter fit for $\tilde{R}_\mathrm{Au}$ and
$\tilde{d}_\mathrm{Au}$, using $\sqrt{B}\eq0.473$\,fm in Eq.~(\ref{eq1}) (\ie the
value corresponding to the lowest collision energy in Table~\ref{T1}, keeping in 
mind that the parameters (\ref{eq14}) are the result of low-energy scattering 
experiments). We thus find
\begin{equation}
\label{eq15}
  \tilde\rho_0\eq0.1695\,\mathrm{fm}^{-3},\ 
  \tilde{R}_\mathrm{Au}\eq6.42\,\mathrm{fm},\ 
  \tilde{d}_\mathrm{Au}\eq0.45\,\mathrm{fm}.
\end{equation}
These values are very close to those obtained in \cite{Hirano:2009ah} for a box-like
nucleon density profile. Fig.~\ref{F1} shows the nucleon center distribution 
$\tilde\rho_\mathrm{Au}$ (Eq.~(\ref{eq13})) and the folded distribution 
$\rho_\mathrm{Au}$ (Eq.~(\ref{eq11})), and compares the latter to a Woods-Saxon 
profile (\ref{eq12}) with the measured parameters (\ref{eq13}).

\section{Energy dependence of initial-state fluctuations in Au+Au collisions} 
\label{sec4}

The smooth dashed curve in Fig.~\ref{F1} can be interpreted as the (unnormalized)
probability (\ie the squared modulus of the single-nucleon wave function) for  finding a 
nucleon centered at position $r$ inside a gold nucleus. At high energies, where the
time for the colliding nuclei to pass through each other is very short, each nucleus-nucleus 
collision can be see as performing an instantaneous position measurement of the positions
of all struck nucleons inside the nuclei (\ie of the $A$-particle wave function in position
space, integrated over the positions of those nucleons that pass through unscathed).
With few exceptions (see e.g. \cite{Alvioli:2009ab}), Monte Carlo implementations of 
these ideas assume uncorrelated nucleons (except for, in some cases, an excluded 
volume corresponding to an infinitely strong hard-core repulsion at short distances), 
\ie a factorization of the $A$-nucleon probability density into single-nucleon probabilities, 
given by the normalized version of Eq.~(\ref{eq13}). This allows to calculate the 
probabilities for distributing the $A$ nucleons inside each nucleus independently.
 
\begin{figure}[h!]
\includegraphics[width=\linewidth]{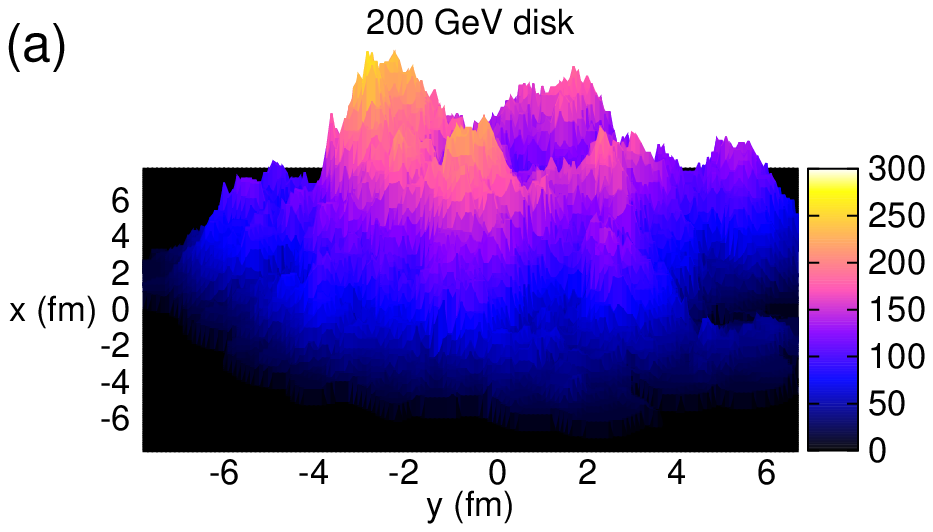}\\
\includegraphics[width=\linewidth]{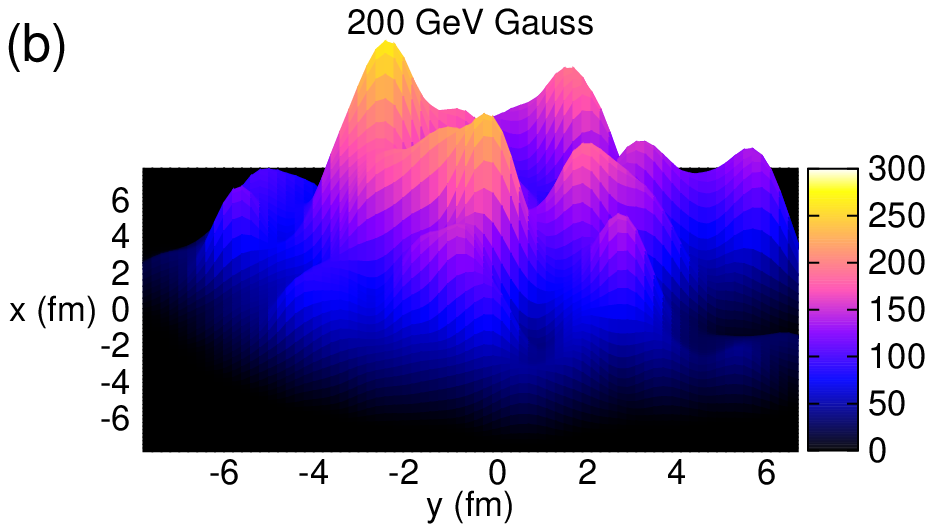}\\
\includegraphics[width=\linewidth]{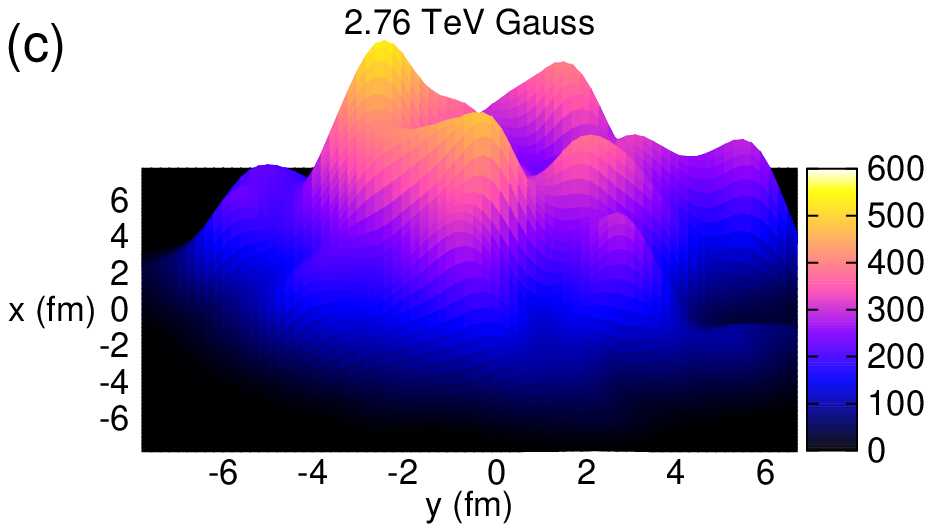}
\caption{(Color online) The transverse entropy density $s(\bm{r}_\perp,\tau_0
  \eq0.4\,\mathrm{fm}/c)$ (in fm$^{-3}$) at mid-rapidity on an 
  $8\,\mathrm{fm}\times8\,\mathrm{fm}$ grid from the MC-KLN model, for 
  a central ($b\eq0$) Au+Au collision. The top graph uses disk-like nucleons (Eq.~(\ref{eq20}) 
  with $\snn\eq42$\,mb), whereas the middle and bottom plots use Gaussian nucleon 
  thickness functions with $\sqrt{B}\eq0.544$ and 0.707\,fm, corresponding to collision 
  energies of $\sqrt{s}=200$\,GeV and 7\,TeV, respectively (see Table~\ref{T1}). All three plots
  use the same collision event, \ie identical spatial distributions for the nucleons in the
  two nuclei.
  \label{F2}
  }
\end{figure}
 
We here use the MC-KLN model described in Ref.~\cite{Drescher:2006ca}. It calculates
initial gluon production by folding the unintegrated gluon distributions of the two nuclei
with a cross section for gluon fusion \cite{Gribov:1984tu} (see 
Refs.~\cite{KLN,Hirano:2005xf,Drescher:2006ca,Hirano:2009ah} for details). The 
unintegrated gluon distributions depend on transverse position through the gluon 
saturation momentum \cite{Drescher:2006ca,Hirano:2009ah}
\begin{equation}
\label{eq16}
  Q^{2}_{s,A}(x,\rperp)= (2\,\mbox{GeV}^{2}) \left(\frac{T_{A}(\rperp)}{T_{A,0}}\right)
                                                                               \left(\frac{0.01}{x}\right)^{\!\lambda}\!\!, 
\end{equation}
where
\begin{equation}
\label{eq17}
  T_A(\rperp) = \int dz\, \rho_A(\rperp,z)
\end{equation}
is the nuclear thickness function of the corresponding nucleus at the same position,
$T_{A,0}\eq1.53$\,fm$^{-2}$, and $x\eq{p}_T\exp(\pm y)/\sqrt{s}$ are the longitudinal
momentum fractions of the gluons from the two colliding nuclei that fuse into
a produced gluon with rapidity $y$ and transverse momentum $p_T$.  The Monte 
Carlo sampling of the position distribution $\tilde\rho_A(\br_0)$ in Eq.~(\ref{eq11}) 
leads to event-by-event fluctuations in the nuclear thickness function $T_A(\rperp)$ 
that (via $Q_s^2(\rperp)$) are reflected in event-by-event fluctuations of the produced 
transverse gluon density which, after thermalization, generates the initial entropy and 
energy density profiles for the hydrodynamic evolution. Through Eq.~(\ref{eq11}), the 
shape and character of these fluctuations is affected by the nucleon density profile 
$\rho_p(r)$.

Integrating Eq.~(\ref{eq11}) over the longitudinal position $z$ yields
\begin{equation}
\label{eq18}
   T_A(\rperp) = \int d^2r_{\perp0} \, \tilde{T}_A(\bm{r}_{\perp0}) \, 
                                                                T_p(\rperp{-}\bm{r}_{\perp0}),
\end{equation}
which, after distributing the nucleons in the nuclei by Monte Carlo sampling the
distribution $\tilde\rho_A(\br_0)$, becomes
\begin{equation}
\label{eq19}
  T_A(\rperp) = \sum_{i=1}^A T_p\left(|\rperp{-}\bm{r}_{i\perp}|\right).
\end{equation}
Here the terms under the sum are Gaussians (\ref{eq2}) with a width $B$ that increases
with $\sqrt{s}$ as described in Sec.~\ref{sec2}. Previous implementations of the MC-KLN
model \cite{Drescher:2006ca,Hirano:2009ah} instead used a disk-like nucleon thickness 
function:
\begin{equation}
\label{eq20}
  T_p(r_\perp) = \frac{\theta(r_\mathrm{N} - r_\perp)}{\snn}, \quad 
  r_\mathrm{N} \equiv \sqrt{\frac{\snn}{\pi}}.
\end{equation}
%

\begin{figure*}[t]
\begin{flushleft}
\begin{tabular}{cc}
\includegraphics[scale=1.23]{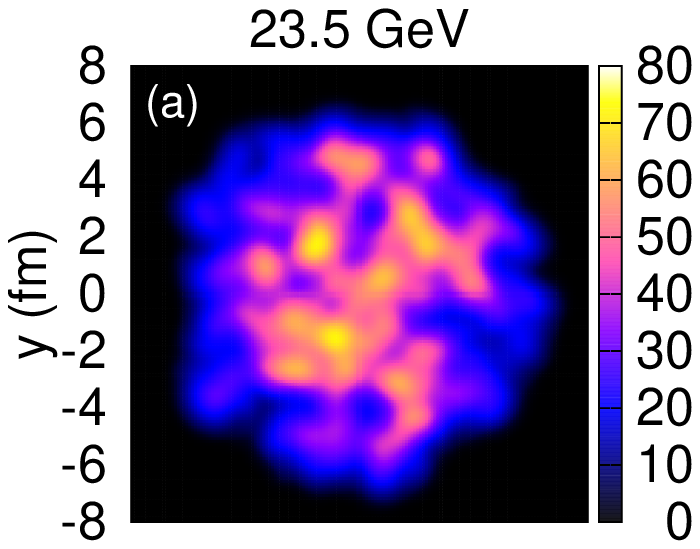}
\includegraphics[scale=1.23]{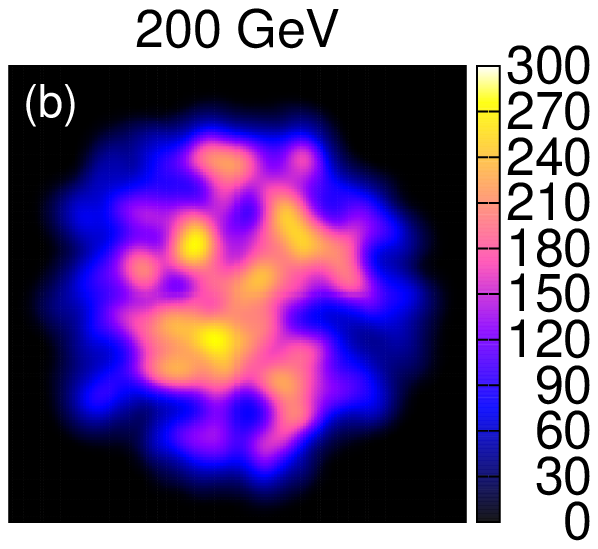}\\
\includegraphics[scale=1.23]{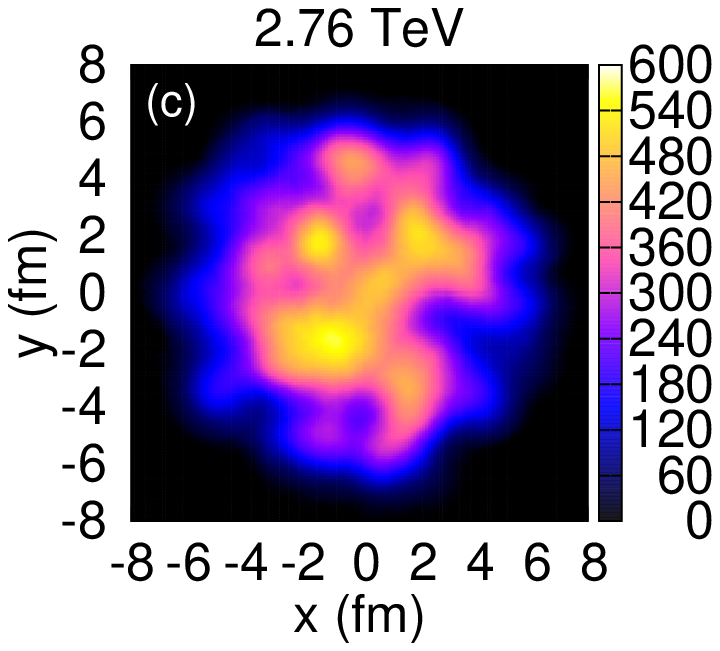}
\includegraphics[scale=1.23]{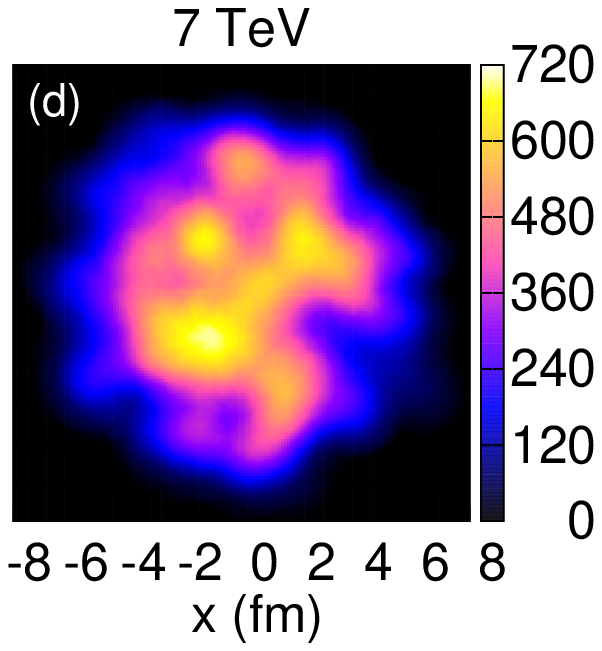}
\end{tabular}
\caption{(Color online) Influence of the growth of the proton with collision energy on 
  the initial transverse entropy density $s(\bm{r}_\perp,\tau_0\eq0.4\,\mathrm{fm}/c)$ 
  (in fm$^{-3}$) for a typical
  central ($b\eq0$) Au+Au collision from the MC-KLN model. All four panels use identical
  nucleon distributions in the two colliding nuclei, but the width $\sqrt{B}$ of the nucleon
  thickness function changes with $\sqrt{s}$ (indicated above each panel) as listed in 
  Table~\ref{T1}.
  \label{F3}
 }
\end{flushleft}
\end{figure*}

In Figs.~\ref{F2} and \ref{F3} we compare the initial transverse gluon density distributions
for a central Au+Au collision with disk-like and Gaussian proton thickness functions and 
for different collision energies, taking into account the growth of the width $\sqrt{B}$ of the 
nucleon thickness function with increasing collision energy. The upper two panels in 
Fig.~\ref{F2} illustrate the significant smoothing of the initial fireball density distribution 
at small length scales that is caused by replacing the disk-like nucleons with Gaussians. 
The small but pervasive discontinuities in the density profile using disk-like nucleons are 
unphysical but constitute a major technical stumbling block for viscous hydrodynamics
which cannot run stably with such discontinuous initial profiles (unless an additional
smoothing step, involving another unphysical parameter, is applied first). With Gaussian
nucleons these discontinuities are naturally washed out.

However, the initial density distributions shown in the middle panel of Fig.~\ref{F2} and in 
the upper two panels of Fig.~\ref{F3} still feature large (physical) inhomogeneities and 
density gradients which, for very early starting times of the hydrodynamic evolution 
(where their effect on the viscous pressure is enhanced by a factor $1/\tau$ 
\cite{Teaney:2003kp} that arises from the longitudinal Bjorken expansion), can drive 
viscous hydrodynamics outside its region of applicability. The length scale of these initial 
state density fluctuations thus generates a lower limit for the hydrodynamic starting time 
$\tau_0$. At earlier times, even if the system started out in a state of local equilibrium, 
viscous forces would drive the system so far away from thermal equilibrium that the 
macroscopic hydrodynamic framework breaks down.

The lower two panels in Fig.~\ref{F2}, as well as the profiles shown in Fig.~\ref{F3},
demonstrate that the growing nucleon size smoothes the initial state density
fluctuations and significantly increases the length scale over which the initial 
pressure profile fluctuates. This implies that for heavy-ion collisions at the LHC viscous hydrodynamics can be applied starting from significantly earlier times than at RHIC 
energies.

\section{Further discussion and conclusions}
\label{sec5}

The smoothing effects of a larger effective nucleon size at higher energies also 
influence the anisotropic flow generated in heavy ion collisions. Initial-state 
density fluctuations entail event-by-event shape fluctuations for the initially
produced fireball. As discussed and studied in detail in 
Refs.~\cite{Alver:2010gr,Qin:2010pf,Qiu:2011iv,Bhalerao:2011bp}, these can be 
characterized by a set of harmonic eccentricity coefficients $\ecc_n$ ($n\eq1,2,\dots$) 
which drive higher order harmonic flow coefficients $v_n$.  

\begin{figure}[h!]
  \includegraphics[width=\linewidth]{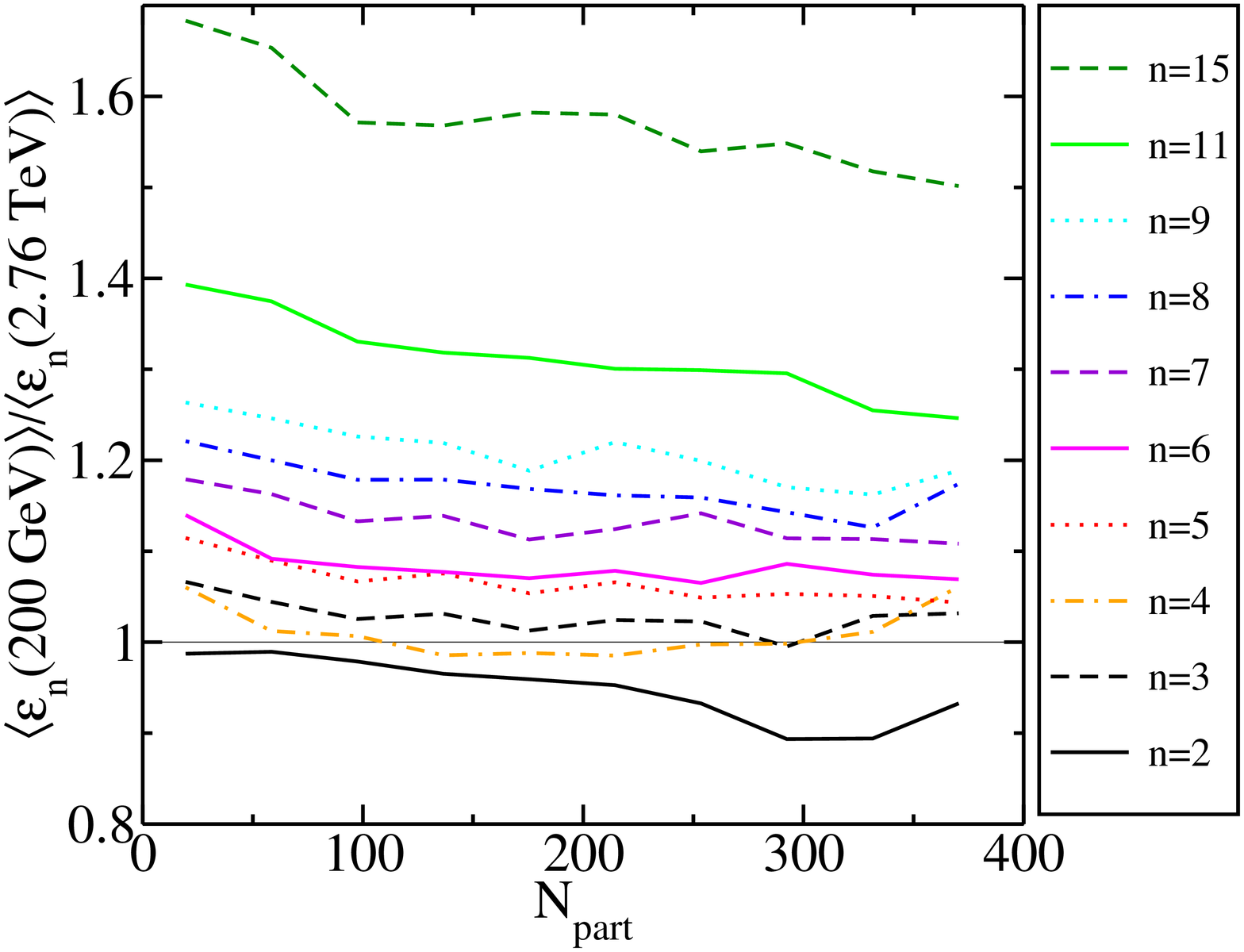}
  \caption{(Color online) 
  The ratio $\La\ecc_n(200\mathrm{GeV})\Ra / \La\ecc_n(2.76\mathrm{TeV})\Ra$ of the 
  ensemble-averaged harmonic eccentricity coefficients $\La\ecc_n\Ra$ ($n\eq2,\dots,15$) 
  at $\sqrt{s}\eq0.2$ and 2.76\,$A$\,TeV, for Au+Au collisions with MC-KLN initial conditions 
  as a function of the number of participating (struck) nucleons, $\Npart$. The plot is based 
  on 95,000 events, binned into 10 equal size $\Npart$ bins containing 9,500 events each.    
  Smoother curves require higher statistics.
  \label{F4}
  }
\end{figure}

Figure~\ref{F4} shows the ratio of the ensemble averages $\La\ecc_n\Ra$ of some of the 
first 15 harmonic eccentricity coefficients at two different energies. The $\ecc_n$ are defined 
in terms of the initial transverse energy density profile $e(r_\perp,\phi)$ through 
\cite{Alver:2010gr,Alver:2010dn}
\begin{equation}
\label{eq21}
  \ecc_n e^{in\psi_n} = - \frac{\int d^2r_\perp\, r_\perp^2\, e^{in\phi}\, e(r_\perp,\phi)} 
                                                   {\int d^2r_\perp\, r_\perp^2\, e(r_\perp,\phi)}.
\end{equation}
Fig.~\ref{F4} is for Au+Au collisions at RHIC and LHC energies as a function of collision 
centrality, using the MC-KLN model with Gaussian nucleons whose width grows with energy. 
We see that, except for the ellipticity $\ecc_2$, all higher order harmonic eccentricity
coefficients are larger at RHIC than at LHC, with the ratio between the two energies
increasing with harmonic order $n$ to about $1.5{-}1.6$ for $n\eq15$. This is a straightforward 
reflection of the higher degree of granularity in the initial density profiles at RHIC than 
at LHC, caused by the smaller effective nucleon size. The ellipticity $\ecc_2$ bucks this 
trend: in contrast to the higher order harmonics which are driven by fluctuations, the 
ellipticity is dominated in all but the most central collisions by the elliptic geometric 
deformation of the nuclear overlap zone. The geometric contribution to the ellipticity 
actually {\it grows} with collision energy, due to the slight swelling of the entire nucleus 
caused by the growth of its constituent nucleons. Consequences for the extraction of the 
QGP shear viscosity from anisotropic flow coefficients will be explored in future work. 

In summary, when taking into account the swelling of the nucleons at higher collision 
energy caused by gluon saturation effects, we expect the elliptic flow to slightly increase,
but the higher order flow harmonics to decrease at the LHC compared to previous 
predictions. This effect increases with the harmonic order of the flow. Furthermore,
smoother event-by-event initial conditions extend the range of validity of a viscous
hydrodynamic description to earlier times, so taking into account the swelling of the
nucleon from RHIC to LHC energies further adds to the applicability of viscous 
hydrodynamics for the description of the dynamical evolution of ultra-relativistic 
heavy-ion collisions.

\acknowledgments{We gratefully acknowledge clarifying discussions with Adrian Dumitru, 
Tetsu Hirano, Will Horowitz, Yuri Kovchegov, Mike Lisa, Yasushi Nara, Zhi Qiu, and Chun 
Shen, and thank Genya Levin for pointing us to Refs.~\cite{Gotsman:2010nw,TOTEM}. JSM
wishes to thank Pasi Huovinen and Dirk Rischke and the Institute for Theoretical Physics
at the J.~W.~Goethe Universit\"at Frankfurt, where this work began, for their kind
hospitality. The work of UH was supported by the U.S.\ Department of Energy 
under Grant No. \rm{DE-SC0004286}. JSM received support from the ExtreMe Matter 
Institute (EMMI) and the BMBF under contract No. 06FY9092, and from the Undergraduate Research Office at The Ohio State University.}


\end{document}